\documentstyle[epsf,eqsecnum,floats,preprint,aps]{revtex}

\tighten

\input epsf
\begin{document}
\newcommand{\beq}{\begin{equation}}
\newcommand{\eeq}{\end{equation}}

\newcommand{\be}{\begin{equation}}
\newcommand{\ee}{\end{equation}}
\newcommand{\bea}{\begin{eqnarray}}
\newcommand{\eea}{\end{eqnarray}}
\newcommand{\PSbox}[3]{\mbox{\rule{0in}{#3}\includegraphics{#1}\hspace{#2}}}
\overfullrule=0pt
\def\Int{\int_{r_H}^\infty}
\def\d{\partial}
\def\e{\epsilon}
\def\M{{\cal M}}
\def\high{\vphantom{\Biggl(}\displaystyle}
\catcode`@=11
\def\@versim#1#2{\lower.7\p@\vbox{\baselineskip\z@skip\lineskip-.5\p@
    \ialign{$\m@th#1\hfil##\hfil$\crcr#2\crcr\sim\crcr}}}
\def\simge{\mathrel{\mathpalette\@versim>}} %
\def\simle{\mathrel{\mathpalette\@versim<}} %
\catcode`@=12 

\def\pr#1#2#3#4{Phys. Rev. D {\bf #1}, #2 (19#3#4)}
\def\prl#1#2#3#4{Phys. Rev. Lett. {\bf #1}, #2 (19#3#4)}
\def\prold#1#2#3#4{Phys. Rev. {\bf #1}, #2 (19#3#4)}
\def\np#1#2#3#4{Nucl. Phys. {\bf B#1}, #2 (19#3#4)}
\def\pl#1#2#3#4{Phys. Lett. {\bf #1B}, #2 (19#3#4)}

\rightline{CU-TP-969}
\rightline{gr-qc/0004001}
\vskip 1cm

\begin{center}
\Large{\bf Monopoles and 
the Emergence of Black Hole Entropy}
\ \\
\ \\
\ \\
\ \\
\ \\
\normalsize{Arthur Lue\footnote{\tt lue@phys.columbia.edu} and 
Erick J. Weinberg\footnote{\tt ejw@phys.columbia.edu}}
\ \\
\ \\
\small{\em Department of Physics \\
Columbia University \\
538 West 120 Street \\
New York, NY 10027}

\end{center}

\begin{abstract}

\baselineskip 16pt
\noindent
One of the remarkable features of black holes is that they possess a
thermodynamic description, even though they do not appear to be
statistical systems.  We use self-gravitating magnetic monopole
solutions as tools for understanding the emergence of this description
as one goes from an ordinary spacetime to one containing a black hole.
We describe how causally distinct regions emerge as a monopole
solution develops a horizon.  We define an entropy that is naturally
associated with these regions and that has a clear connection with the
Hawking-Bekenstein entropy in the critical black hole limit.

\end{abstract}

\setcounter{page}{0}
\thispagestyle{empty}
\maketitle

\eject

\vfill

\baselineskip 18pt plus 2pt minus 2pt

Black holes have long captured the modern imagination.  These objects,
containing spacetime singularities hidden behind event horizons,
manifest features both striking and surprising.  Among these is the
fact that thermodynamic properties can be ascribed to black holes,
even though they do not appear to be statistical systems.  In this
essay, we discuss how a thermodynamic description emerges as one goes
from a normal spacetime to a spacetime containing a black hole.

How can one investigate the transition from a nonsingular spacetime to
one with a horizon?  Stars and other astrophysical sources that
collapse and form event horizons when they exceed a critical density
and size offer one possible direction.  However, the onset of black
hole behavior happens when the horizon is infinitesimally small, with
infinite curvatures; quantum effects that are presumably important
here are as yet poorly understood.

Self-gravitating magnetic monopoles \cite{orig1,ortiz,orig2,orig2b,LW}
offer another class of laboratories for investigating the onset of
black-hole behavior.  They have the advantage of being parametrically
tunable systems in which the approach to the black hole limit can be
implemented by increasing the soliton mass scale relative to the
Planck mass \cite{LW2}.  Furthermore, by appropriate choice of
parameters one can make the horizon radius of the critical solution 
arbitrarily large and the curvatures arbitrarily small, thus ensuring
that quantum gravity effects can be safely ignored.

\subsection{Self-gravitating magnetic monopoles}

For our purposes it is sufficient to consider spherically symmetric
spacetimes, for which the metric can be written in the form
$$
     ds^2 = B dt^2 - A dr^2 - r^2 (d\theta^2 + \sin^2 \theta\,
     d\phi^2)\ .
$$
In general, a horizon corresponds to a zero of $1/A$; the horizon is
extremal if $d(1/A)/dr$ also vanishes.  We work in the context of an
SU(2) gauge theory with gauge coupling $e$ and a triplet Higgs field
whose vacuum expectation value $v$ breaks the symmetry down to
U(1).  The elementary particle spectrum of the theory includes a
neutral massive Higgs particle, a pair of electrically charged vector
bosons, and a massless photon.

In flat spacetime this theory possesses a finite energy monopole
solution with magnetic charge $Q_M=4\pi/e$ and mass $M\sim v/e$.  This
monopole has a core region, of radius $\sim 1/ev$, in which there are
nontrivial massive fields.  Beyond this core is a Coulomb region in
which the massive fields rapidly approach their vacuum values, leaving
only the Coulomb magnetic field.

The effects of adding gravitational interactions depend on the value of
$v$.  For $v$ much less than the Planck mass $M_{\rm Pl}$, one finds
that $1/A$ is equal to unity at the origin, decreases to a minimum at
a radius of order $1/ev$, and then increases again with $A(\infty)
=1$.  As $v$ is increased, this minimum becomes deeper, until an
extremal horizon develops at a critical value $v_{\rm cr}$ of the
order $M_{\rm Pl}$; interestingly, the interior remains nonsingular.
We will refer to the radius, $r=r_*$, at which
$1/A = (1/A)_{\rm min}$ as the quasi-horizon.

\subsection{Probing the quasi-black hole}

\begin{figure} \begin{center}\PSbox{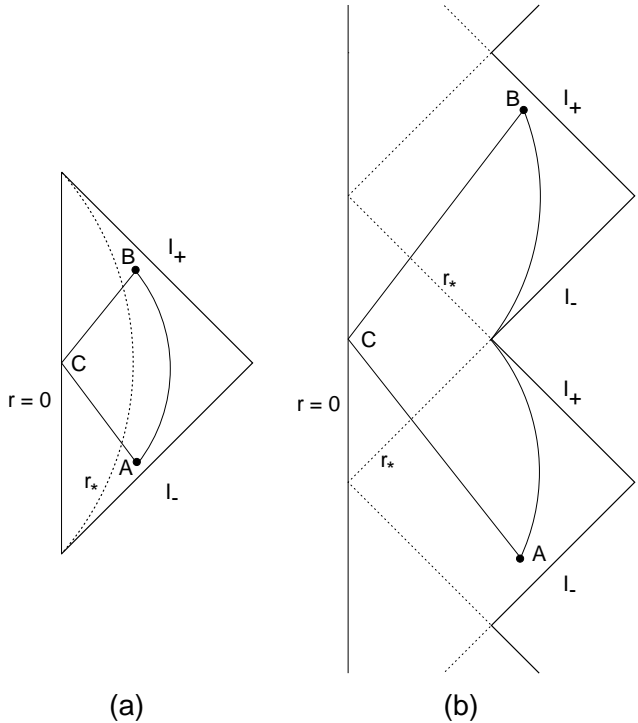
hscale=50 vscale=50 hoffset=50 voffset=0}{4in}{3.5in}\end{center}
\caption{
Penrose diagrams for (a) subcritical monopole and (b) critical
monopole black hole.  In the former case $r_*$ represents the
quasi-horizon whereas in the latter case that radius represents
a true horizon.
}
\label{fig:penrose}
\end{figure}

For any $v<v_{\rm cr}$, the self-gravitating monopole solution is a
nonsingular spacetime with a Penrose diagram of the same form as that
of Minkowski spacetime (Fig.~\ref{fig:penrose}a).  The critical
solution, on the other hand, can be extended beyond the original
coordinate patch to yield a spacetime with the Penrose diagram shown
in Fig.~\ref{fig:penrose}b.  This diagram is quite similar to that of
an extremal RN black hole, but differs from it by not having a
singularity at $r=0$.  The difference between the two diagrams is
striking and seems to indicate a discontinuity at $v=v_{\rm cr}$, in
contradiction with the usual expectation that physical quantities
should vary continuously with the parameters of a theory.  However,
this discontinuity is perhaps better viewed as an artifact of the
conformal transformation that produces the Penrose diagram from an
infinite spacetime.  This can be seen by considering an observer who
remains at a radius $r=r_{\rm obs} \gg r_*$ and probes the
interior of the quasi-black hole by sending in a particle along the
trajectory ACB shown in Fig.~\ref{fig:penrose}a.  As the probe moves
along this trajectory, the elapsed coordinate time (which is
approximately the same as the elapsed proper time of the observer) is
$$
    \Delta t = 2 \int_0^{r_{\rm obs}} dr\, {dt/d\tau \over
                   dr/d\tau} 
       = 2 \int_0^{r_{\rm obs}} dr\,
     {A \over \sqrt{AB}} \left[ 1 -{B \over E^2} \left({J^2 \over r^2}
      + 1\right) \right]^{-1/2}
$$
where $E$ is the probe's energy and $J$ is its angular momentum.

Consistency with our physical expectations of continuity requires that
$\Delta t$ diverge as the quasi-black hole approaches the critical
limit and $\e =(1/A)_{\rm min} \rightarrow 0$.  If this happens, then
the region containing point B would become effectively disconnected from
that containing point A, just as in the black hole Penrose diagram of 
Fig.~\ref{fig:penrose}b.  By examining the behavior of the metric
functions as the quasi-black hole approaches the critical limit, we
find that in this limit
$$
        \Delta t \approx k\e^{-q}   + \cdots
$$
where the exponent $q$ depends on specific monopole parameters
but is always greater than or equal to $0.5$.  A similar result is
obtained if one considers probing the black hole interior by sending
in waves of some classical field.  Thus, the time needed for an
external observer to obtain information from the interior region
diverges in the critical limit.  Most importantly, the leading
contribution to $\Delta t$ is determined solely by the spacetime
geometry and is independent of the energy, angular momentum, or other
features of the probe.

\subsection{Entropy and thermodynamics}

Until a horizon is actually formed, the interior of the quasi-black
hole can be probed by external observers of infinite patience and
lifetime.  However, for an observer with a finite lifetime $T$, the
interior region of a near-critical configuration becomes inaccessible
once $\e \lesssim T^{-1/q}$.  Such an observer would most naturally
describe any larger system containing this configuration in terms of a
density matrix $\rho$ obtained by tracing over the degrees of freedom
inside the quasi-horizon.  Using this density matrix one can define an
entropy $S_{\rm interior}= - {\rm Tr}\,\rho \ln \rho$ that can be
associated with the interior of the quasi-black hole.

One could, of course, proceed in this manner to define an entropy for
any arbitrary region in space, just as one can choose to make the
information in any subsystem inaccessible by putting the subsystem
behind a locked door.  The crucial difference here is that the
inaccessibility is due to the intrinsic properties of the spacetime,
and that the boundary of the inaccessible region is defined by the system
itself rather than by some arbitrary external choice.  It is thus
reasonable to define $S_{\rm interior}$ as {\it the} entropy of the
quasi-black hole.

A precise calculation of this entropy is clearly infeasible.  Among
other problems, such a calculation would require a correct
implementation of an ultraviolet cutoff, which presumably would
require a detailed understanding of how to perform the calculation in
the context of a consistent theory of quantum gravity.  As an initial
effort, one can take the ultraviolet cutoff to be the Planck mass
$M_{\rm Pl}$ and try to obtain an order of magnitude estimate.  To see
what result might be expected, we recall a calculation of Srednicki
\cite{srednicki}, who showed that tracing over the degrees of freedom
of a scalar field inside a region of flat spacetime with surface area
$A$ leads to an entropy $S=\kappa M^2 A$, where $M$ is the ultraviolet
cutoff and $\kappa$ is a numerical constant.  Although the value of
$\kappa$ depends on the details of the theory, general arguments
\cite{srednicki} suggest that an entropy obtained in this fashion
should always be proportional to the surface area.  Hence, we expect
that the entropy associated with our quasi-black hole is $S_{\rm
interior} \sim M_{\rm Pl}^2 A$.  A very plausible guess is that in the
critical limit this goes precisely to the standard black hole result
$S_{\rm BH} = M_{\rm Pl}^2 A/4$.

The suggestion that the entropy of a black hole might be understood in
terms of the degrees of freedom inside the horizon is not a new idea.
However, any attempt to make this idea more precise must overcome the
difficulties that the ``interior'' region of a black hole is not
static and that it contains a singularity.  In contrast, our spacetime
configurations are static and topologically trivial.  Because their
interiors can be unambiguously defined, it is conceptually clear what
it means to trace over the interior degrees of freedom, even though it
may not yet be possible to implement this calculation in complete
detail.

Our calculations suggest an understanding of how a thermodynamics
description emerges as one moves from a flat space configuration to a
black hole.  In any thermodynamic description of a system there is an
implicit time scale $\tau$ that separates fast processes accounted for
in the thermodynamics from slow process that are not.  One assumes
that for times shorter than $\tau$ the system can be described as
effectively in equilibrium; for true equilibrium, this time scale is
infinite.  Correspondingly, self-gravitating monopoles have an
associated time scale that gives the minimum time needed for an
external observer to probe the interior; on shorter time scales, the
monopole is effectively a statistical system as far as the observer is
concerned.  In the limit where the horizon forms and the monopole
becomes a true black hole, this time scale becomes truly infinite and
the thermodynamic description becomes exact.

\acknowledgments

This work was supported in part by the U.S. Department of Energy.


\begin{thebibliography}{}

\bibitem{orig1}
K. Lee, V. P. Nair, and E. J. Weinberg, Phys. Rev. D
{\bf 45}, 2751 (1992).

\bibitem{ortiz}
M. E. Ortiz, \pr{45}{R2586}92.

\bibitem{orig2}
P. Breitenlohner, P. Forg\'acs, and D. Maison,
Nucl. Phys. {\bf B383}, 357 (1992);

\bibitem{orig2b}
P. Breitenlohner, P. Forg\'acs, and D. Maison,
Nucl. Phys. {\bf B442}, 126 (1995).

\bibitem{LW}
A. Lue and E. J. Weinberg, Phys. Rev. D {\bf 60}, 084025.

\bibitem{LW2}
A. Lue and E. J. Weinberg, hep-th/0001140, to appear in Physical
Review D.  

\bibitem{srednicki}
M. Srednicki, Phys. Rev. Lett. {\bf 71}, 666 (1993).
\end{thebibliography}
\end{document}